\documentclass{llncs}
\usepackage{makeidx}  
\usepackage[nolist]{acronym}
\usepackage{textcomp} 
\usepackage{color}
\usepackage{layouts}
\usepackage{booktabs}
\usepackage{mathptmx}
\usepackage{graphicx}
\usepackage{hyperref}
\usepackage{breakurl}

\begin{document}
\begin{acronym}[DVFS]
    \acro{AGU}{address generation unit}
    \acro{AVX}{advanced vector extensions}
    \acro{CA}{cache agent}
    \acro{CL}{cache line}
    \acro{CoD}{cluster-on-die}
    \acro{DCT}{dynamic concurrency throttling}
    \acro{DIR}{directory}
    \acro{DP}{double precision}
    \acro{DP}{double precision}
    \acro{DVFS}{dynamic voltage and frequency scaling}
    \acro{ECM}{execution-cache-memory}
    \acro{ES}{early snoop}
    \acro{FMA}{fused multiply-add}
    \acro{FP}{floating-point}
    \acro{HA}{home agent}
    \acro{HS}{home snoop}
    \acro{IMCI}{initial many core instructions}
    \acro{LFB}{line fill buffer}
    \acro{LLC}{last-level cache}
    \acro{MC}{memory controler}
    \acro{MSR}{model specific register}
    \acro{NT}{non-temporal}
    \acro{NUMA}{non-uniform memory access}
    \acro{OSB}{opportunistic snoop broadcast}
    \acro{RAPL}{running average power limit}
    \acro{SIMD}{single instruction, multiple data}
    \acro{SKU}{stock keeping unit}
    \acro{SP}{single precision}
    \acro{SP}{single precision}
    \acro{SSE}{streaming SIMD extensions}
    \acro{TDP}{thermal design power}
    \acro{UFS}{Uncore frequency scaling}
\end{acronym}
\frontmatter          
\pagestyle{headings}  
\addtocmark{Hamiltonian Mechanics} 

\mainmatter              
\title{An analysis of core- and chip-level architectural features in four generations of Intel server processors}
\author{}
\institute{}
%
\titlerunning{Performance on Intel processors}  
%
\author{Johannes Hofmann\inst{1} \and Georg Hager\inst{2} \and Gerhard Wellein\inst{2} \and Dietmar Fey\inst{1}}
\authorrunning{Johannes Hofmann et al.} 
%
\tocauthor{Johannes Hofmann, Georg Hager}
\institute{Computer Architecture, University of Erlangen-Nuremberg, 91058 Erlangen, Germany,\\
\email{johannes.hofmann@fau.de, dietmar.fey@fau.de}
\and
Erlangen Regional Computing Center (RRZE), 91058 Erlangen, Germany,\\
\email{georg.hager@fau.de, gerhard.wellein@fau.de}}

\maketitle              

\vspace*{-1cm}

\begin{abstract}
This paper presents a survey of architectural features among four
generations of Intel server processors (Sandy Bridge, Ivy Bridge,
Haswell, and Broadwell) with a focus on performance with floating
point workloads. Starting on the core level and going down the memory
hierarchy we cover instruction throughput for floating-point
instructions, L1 cache, address generation capabilities, core clock
speed and its limitations, L2 and L3 cache bandwidth and latency, the
impact of Cluster on Die (CoD) and cache snoop modes, and the Uncore
clock speed. Using microbenchmarks we study the influence of these
factors on code performance. This insight can then serve as input for
analytic performance models. We show that the energy efficiency of the
LINPACK and HPCG benchmarks can be improved considerably by tuning the
Uncore clock speed without sacrificing performance, and that the
Graph500 benchmark performance may profit from a suitable choice of
cache snoop mode settings.

\end{abstract}

\section{Introduction}

Intel Xeon server CPUs dominate in the commodity HPC market. Although
the microarchitecture of those processors is ubiquitous and can also
be found in mobile and desktop devices, the average developer of
numerical software hardly cares about architectural details and relies
on the compiler to produce ``decent'' code with ``good''
performance. If we actually want to know what ``good performance''
means we have to build analytic models that describe the
interaction between software and hardware. Despite the necessary
simplifications, such models can give useful hints towards the
relevant bottlenecks of code execution and thus point to 
viable optimization approaches. The
Roof{}line model \cite{hockney89,roofline:2009} and 
the Execution-Cache-Memory (ECM) model \cite{Hager:2012,sthw15}
are typical examples. Analytic modeling requires
simplified machine and execution models, with details about
properties of execution units, caches, memory, etc.
Although much of this data is provided by manufacturers, many
relevant features can only be understood via microbenchmarks,
either because they are not documented or because the hardware
cannot leverage its full potential in practice. One simple
example is the maximum memory bandwidth of a chip, which can be calculated
from the number, frequency, and width of the DRAM channels
but which, in practice, may be significantly lower than this
absolute limit. Hence, microbenchmarks such as STREAM \cite{McCalpin:1995}
or \verb.likwid-bench. \cite{Treibig:2011:3} are used to measure
the limits achievable in practice.

Although there has been some convergence in processor microarchitecures
for high performance computing, the latest CPU models show
interesting differences in their performance-relevant
features. Building good analytic performance models and,
in general, making sense of performance data, requires
intimate knowledge of such details. The main goal of this paper
is to provide a coverage and critical discussion of those
details on the latest four Intel architecture generations
for server CPUs: Sandy Bridge (SNB), Ivy Bridge (IVB), Haswell (HSW),
and Broadwell (BDW)\@. The actual CPU models used for the 
analysis are described in Sect.~\ref{sec:hardware} below.

\subsection{Performance on modern multicore CPUs}

Out of the many possible approaches to performance analysis and
optimization (coined \emph{performance engineering} [PE]) we favor
concepts based on analytic performance models. For recent server
multicore designs the ECM performance model allows for a very accurate
description of single-core performance and scalability. In contrast to
the Roof{}line model it drops the assumption of a single bottleneck
for the steady-state execution of a loop.  Instead, time contributions
from in-core execution and data transfers through the memory hierarchy
are calculated and then put together according to the properties of a
particular processor architecture; for instance, in Intel x86 server CPUs
all time contributions from data transfers including LOADs and STOREs
in the L1 cache must be added to get a
prediction of single-core data transfer time \cite{sthw15,Hofmann2016}\@. On the
other hand, the IBM Power8 processor shows almost perfect overlap
\cite{Hofmann:2016-Kahan}\@.  A full introduction to the 
ECM model would exceed the scope of this paper, so we refer to the 
references given above. The model has been shown to work well
for the analysis of implementations of several important computational 
kernels \cite{Treibig:2012:2,sthw15,Wittmann:2016,Gasc:2016,Hofmann:2016-Kahan}\@.

In order to construct analytic models
accurately, data about the capabilities of the microarchitecture and
how it interacts with the code at hand is needed. For floating-point
centric code in scientific computing, maximum throughput and latency
numbers for arithmetic and LOAD/STORE instructions
are most useful in all their vectorized and non-vectorized, single (SP) and
double precision (DP) variants.  On Intel multicore CPUs up to Haswell,
this encompasses scalar, \ac{SSE}, \ac{AVX}, and AVX2 instructions.  Modeling
the memory hierarchy in the ECM model requires the maximum data
bandwidth between adjacent cache levels (assuming that the hierarchy
is inclusive) and the maximum (saturated) memory bandwidth. As for the
caches it is usually sufficient to assume the maximum documented
theoretical bandwidth (presupposing that all prefetchers work
perfectly to hide latencies), although latency penalties might apply
\cite{Hofmann:2016-Kahan}\@. The main memory bandwidth and latency may
depend on the \ac{CoD} mode and cache snoop mode
settings. Finally, the latest Intel CPUs work with at least two clock
speed domains: one for the core (or even individual cores) and one for
the Uncore, which includes the L3 cache and memory controllers.  Both
are subject to automatic changes; in case of \ac{AVX}
code on Haswell and later CPUs the guaranteed baseline clock speed is
lower than the standard speed rating of the chip.  The
performance and energy consumption of code depends crucially on the
interplay between these clock speed domains. Finally, especially
when it comes to power dissipation and capping, considerable variations
among the specimen of the same CPU model can be observed.

All these intricate architectural details influence benchmark
and application performance, and it is insufficient to look up the
raw specs in a data sheet in order to understand this influence.

\subsection{Related work}

There is a large number of papers dealing with details in the
architecture of CPUs and their impact on performance and energy
consumption. In \cite{Barker:2008} the authors assessed the
capabilities of the then-new Nehalem server processor for workloads in
scientific computing and compared its capabilities with its
predecessors and competing designs. In \cite{Schoene:2014}, tools and
techniques for measuring and tuning power and energy consumption of
HPC systems were discussed. The QuickPath Interconnect (QPI) snoop
modes on the Haswell EP processor were investigated in
\cite{Molka:2015}\@.  Energy efficiency features, including the AVX
and Uncore clock speeds, on the same architecture were studied in
\cite{Hackenberg:2015} and \cite{Hofmann:2016:E2SC}\@. Our work
differs from all those by systematically investigating relevant
architectural features, from the core level down to memory,
via microbenchmarks in view of analytic
performance modeling as well as important benchmark workloads such as
LINPACK, Graph500, and HPCG.

\subsection{Contribution}

Apart from confirming or highlighting some documented or previously
published findings, this paper makes the following new contributions:
\begin{itemize}
\item We present benchmark results showing the improvement in
  the performance of the vector gather instruction from HSW to BDW.
  On BDW it is now advantageous to actually use the gather 
  instruction instead of ``emulating'' it.
\item We fathom the capabilities of the L2 cache on all four
  microarchitectures and establish practical limits for L2 bandwidth
  that can be used in analytic ECM modeling. These limits are
  far below the advertised 64\,B/cy on HSW and BDW.
\item We study the bandwidth scalability of the L3 cache depending
  on the Cluster on Die (CoD)  mode and show that, although the parallel
  efficiency for streaming code is never below 85\%, CoD has 
  a measurable advantage over non-CoD.
\item We present latency data for all caches and main memory under
  various cache snoop modes and CoD/non-CoD. We find that although CoD
  is best for streaming and \ac{NUMA} aware workloads in terms of latency
  and bandwidth, highly irregular, NUMA-unfriendly code such as the
  Graph500 benchmark benefits dramatically from non-CoD mode with Home
  Snoop and Opportunistic Snoop Broadcast by as much as 50\% on BDW.
\item We show how the Uncore clock speed on HSW and BDW has
  considerable impact on the power consumption of bandwidth- and
  cache-bound code, opening new options for energy efficient
  and power-capped execution.
\end{itemize}

\section{Test bed}

\subsection{Hardware description}\label{sec:hardware}

All measurements were performed on standard two-socket Intel Xeon servers. A
summary of key specifications of the four generations of processors is shown in
Table~\ref{tab:testbed}. According to Intel's ``tick-tock'' model, a ``tick''
represents a shrink of the manufacturing process technology; however, it should
be noted that ``ticks'' are often accompanied by minor microarchitectural
improvements while a ``tock'' usually involves larger changes. 
\begin{table*}[!tb]
\centering
\caption{\label{tab:testbed}Key test machine specifications. 
All reported numbers taken from data sheets.}
\resizebox{\textwidth}{!}{%
\begin{tabular}{l c c c c }
\toprule
Microarchitecture               & Sandy Bridge-EP                   & Ivy Bridge-EP                         & Haswell-EP                            & Broadwell-EP                                      \\
\midrule
Shorthand                       & SNB                               & IVB                                   & HSW                                   & BDW                                               \\
Chip Model                      & Xeon E5-2680                      & Xeon E5-2690 v2                       & Xeon E5-2695 v3                       & E5-2697 v4                                        \\
Release Date                    & Q1/2012                           & Q3/2013                               & Q3/2014                               & Q1/2016                                           \\
Base Freq.                      & 2.7\,GHz                          & 3.0\,GHz                              & 2.3\,GHz                              & 2.3\,GHz\\
Max All Core Turbo Freq.        & ---                               & ---                                   & 2.8\,GHz                              & 2.8\,GHz\\
AVX Base Freq.                  & ---                               & ---                                   & 1.9\,GHz                              & 2.0\,GHz\\
AVX All Core Turbo Freq.        & ---                               & ---                                   & 2.6\,GHz                              & 2.7\,GHz\\
Cores/Threads                   & 8/16                              & 10/20                                 & 14/28                                 & 18/36                                             \\
Latest SIMD Extensions          & AVX                               & AVX                                   & AVX2, FMA3                            & AVX2, FMA3                                        \\
Memory Configuration            & 4 ch. DDR3-1600                   & 4 ch. DDR3-1866                       & 4 ch. DDR4-2133                       & 4 ch. DDR4-2400                                   \\
Theor. Mem. Bandwidth           & 51.2\,GB/s                        & 59.7\,GB/s                            & 68.2\,GB/s                            & 76.8\,GB/s                                        \\
L1 Cache Capacity               & 8$\times$32\,kB                   & 10$\times$32\,kB                      & 14$\times$32\,kB                      & 18$\times$32\,kB              \\
L2 Cache Capacity               & 8$\times$256\,kB                  & 10$\times$256\,kB                     & 14$\times$256\,kB                     & 18$\times$256\,kB                \\
L3 Cache Capacity               & 20\,MB (8$\times$2.5\,MB)         & 25\,MB (10$\times$2.5\,MB)            & 35\,MB (14$\times$2.5\,MB)            & 45\,MB (18$\times$2.5\,MB)                        \\
L1$\rightarrow$Reg Bandwidth    & 2$\times$16\,B/cy                 & 2$\times$16\,B/cy                     & 2$\times$32\,B/cy                     & 2$\times$32\,B/cy                                 \\
Reg$\rightarrow$L1 Bandwidth    & 1$\times$16\,B/cy                 & 1$\times$16\,B/cy                     & 1$\times$32\,B/cy                     & 1$\times$32\,B/cy                                 \\
L1$\leftrightarrow$L2 Bandwidth & 32\,B/cy                          & 32\,B/cy                              & 64\,B/cy                              & 64\,B/cy                                          \\
L2$\leftrightarrow$L3 Bandwidth & 32\,B/cy                          & 32\,B/cy                              & 32\,B/cy                              & 32\,B/cy                                          \\
\bottomrule
\end{tabular}
} 
\end{table*}

SNB (a ``tock'') first introduced \ac{AVX}, doubling the \ac{SIMD}
width from \ac{SSE}'s 128\,bit to 256\,bit. One major shortcoming of SNB
is directly related to AVX: Although the \ac{SIMD} register width has
doubled and a second LOAD unit was added, data path widths between the L1 cache and
individual LOAD/STORE units were left at 16\,B/cy.  This leads to AVX stores
requiring two cycles to retire on SNB, and AVX LOADs block both units.  
IVB, a ``tick'', saw an increase in core count as well as a higher memory
clock; in addition, IVB brought speedups for several
instructions, e.g., \ac{FP} divide and square
root; see  Table~\ref{tab:instructions} for details.

HSW, a ``tock'', introduced AVX2, extending the existing 256\,bit SIMD
vectorization from floating-point to integer data types.  Instructions
introduced by the \ac{FMA} extension are handled by two new, AVX-capable
execution units. Data path widths between the L1 cache and registers as well as
the L1 and L2 caches were doubled. A vector gather instruction provides a
simple means to fill SIMD registers with non-contiguous data, making it
easier for the compiler to vectorize code with indirect accesses.
To maintain scalability of
the core interconnect, HSW chips with more than eight cores move from a
single-ring core interconnect to a dual-ring design.  At the same time,
HSW introduced the new \ac{CoD} mode, in which a chip is optionally
partitioned into two equally sized \ac{NUMA} domains in order
to reduce latencies and increase scalability. 
Starting with HSW, the system's QPI
snoop mode can also be configured. 
HSW no longer guarantees to run at the base frequency with AVX code.
The guaranteed frequency when running AVX code on all cores is referred to as
``AVX base frequency,'' which can be significantly lower than the nominal frequency
\cite{Intel:xeon-e5-v3-spec-update,Microway:xeon-e5-v4-specs}.
Also there is a separation of frequency domains between cores and Uncore.
The Uncore clock is  now independent and can either be set
automatically (when \ac{UFS} is enabled) or manually via \acp{MSR}.

As a ``tick,'' BDW, the most recent Xeon-EP processor, offers minor
architectural improvements. Floating-point and gather instruction
latencies and throughput have partially improved.  The dual-ring
design was made symmetric and an additional QPI snoop mode 
is available.



\subsection{Software and benchmarks}

All high-level language benchmarks (Graph500, HPCG) were compiled
using Intel ICC 16.0.3. For Graph500 we used the reference implementation
in version 2.1.4, and for LINPACK we ran the Intel-provided binary
contained in MKL 2017.1.013, the most recent version available at the
time of writing. 

The LIKWID\footnote{\url{http://tiny.cc/LIKWID}} tool suite in its
current stable version 4.1.2 was employed heavily in many of our
experiments.  All low-level benchmarks consisted of hand-written
assembly. When available (e.g., for streaming kernels auch as STREAM
triad and others) we used the assembly implementations in the
\verb.likwid-bench. microbenchmarking tool. Latency measurements in
the memory hierarchy were done with all prefetchers turned off (via
\verb.likwid-features.) and a pointer chasing code that ensures
consecutive cache line accesses. Energy
consumption measurements were taken with the
\verb.likwid-perfctr. tool via the RAPL (Running Average Power Limit)
interface, and the clock speed of the CPUs was controlled with
\verb.likwid-setFrequencies.\@. 


\section{In-core features}

\subsection{Core frequency}
\label{sec:core_frequency}

Starting with HSW, Intel chips offer different base and turbo frequencies for AVX and
SSE or scalar instruction mixes. This is due to the higher power requirement of
using all SIMD lanes in case of AVX. To reflect this behavior,
Intel introduced a new frequency nomenclature for these chips.

The ``base frequency,'' also known as the ``non-AVX base frequency'' or
``nominal frequency'' is the minimum frequency that is guaranteed when running
scalar or \ac{SSE} code on all cores. This is also the frequency the chip is
advertised with, e.g., 2.30\,GHz for the Xeon E5-2695v3 in
Table~\ref{tab:testbed}. The maximum frequency that can be achieved when
running scalar or SSE code on all cores is called ``max all core turbo
frequency.''
The ``AVX base frequency'' is the minimum frequency that is guaranteed when
running AVX code on all cores and is typically significantly lower than the
(non-AVX) base frequency. Analogously, the maximum frequency that can be
attained when running AVX code is called ``AVX max all core turbo frequency.''

On HSW, at least core running AVX code resulted in a chip-wide frequency
restriction to the AVX max all core turbo frequency. On BDW, cores running
scalar or SSE code are allowed to float between the non-AVX base and max all
core turbo frequencies even when other cores are running AVX code.

\begin{figure}[tb]
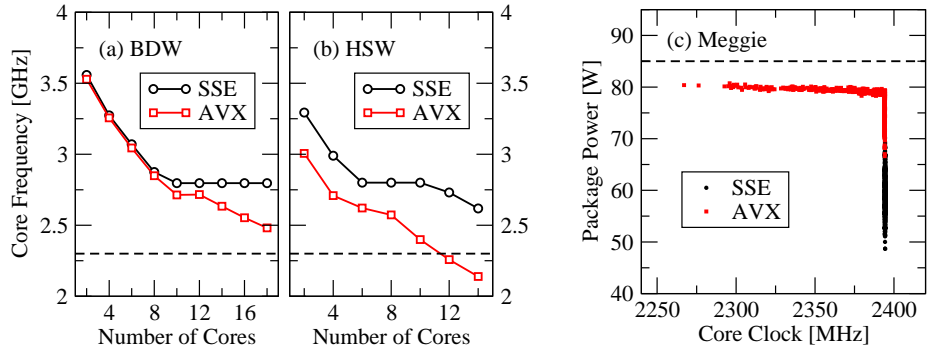

    \centering
    \includegraphics*[scale=0.5]{broadep2-freq} 
    \includegraphics*[scale=0.5]{hasep1-freq}\hfill
    \includegraphics*[scale=0.5]{meggie-freq-tdp}
    \caption{\label{fig:core_freqs}Attained chip frequency during LINPACK runs on all cores on (a) BDW and (b) HSW. (c) Variation of clock speed and package power among all
      1456 Xeon E5-2630v4 CPUs in RRZE's ``Meggie'' cluster running LINPACK.}
\end{figure}

All relevant values for the HSW and BDW specimen used can be found in
Table~\ref{tab:testbed}.  According to official documentation the actually used frequency
depends on the workload; more specifically, it depends on
the percentage of AVX instructions in a certain 
instruction execution window.
To get a better idea about what to expect for demanding workloads,
LINPACK and FIRESTARTER \cite{Hackenberg:2013} were selected to determine
those frequencies. The maximum frequency difference between both benchmarks
was 20\,MHz, so Figure~\ref{fig:core_freqs} shows only results obtained with
LINPACK. Figure~\ref{fig:core_freqs}a shows that BDW can maintain a frequency well
above the AVX \emph{and} the non-AVX base frequency for workloads running at
its TDP limit of 145\,W (measured package power during stress tests was
144.8\,W). HSW, shown in Figure~\ref{fig:core_freqs}b, drops below the non-AVX
base frequency of 2.3\,GHz, but stays well above the AVX base frequency of
1.9\,GHz while consuming 119.4\,W out of a 120\,W TDP. When running SSE
LINPACK, BDW consumes 141.8\,W and manages to run at the max all core turbo
frequency of 2.8\,GHz.  On HSW, running LINPACK with SSE instructions still
keeps the chip at its TDP limit (119.7\,W out of 120\,W); the attained
frequency of 2.6\,GHz is slightly below the max all core turbo frequency of
2.7\,GHz.

While it might be tempting to generalize from these results, we must emphasize
that statistical variations even between specimen of the same CPU type are very
common~\cite{Wilde:2015}. When examining all 1456 Xeon E5-2630v4 (10-core,
2.2\,GHz base frequency) chips of
RRZE's new ``Meggie'' cluster,\footnote{\url{http://www.hpc.rrze.fau.de/systeme/meggie-cluster.shtml}} we found significant variations
across the individual CPUs. The chip has a max all core turbo and AVX max all core turbo
frequency of 2.4\,GHz \cite{Microway:xeon-e5-v4-specs}.
Figure~\ref{fig:core_freqs}c shows each chip's frequency and package power
when running LINPACK with SSE or AVX on all cores.
With SSE code, each chip manages to attain the
max all core turbo frequency of 2.4\,GHz. However, a variation in power
consumption can be observed.  When running AVX code, not all chips 
reach the defined peak frequency but stay well above the AVX base frequency of
1.8 \,GHz. Some chips do hit the frequency ceiling; for these,
a strong variation can be observed in the power domain.

\subsection{Instruction throughput and latency}\label{sec:instructions}

Accurate predictions of instruction execution (i.e., how many clock
cycles it takes to execute a loop body assuming a steady state
situation with all data coming from the L1 cache) are notoriously
difficult in all but the simplest cases, but they are needed as input
for analytic models. As a ``lowest-order'' and most
optimistic approximation one can assume full throughput, i.e., all
instructions can be executed independently and are dynamically fed to
the execution ports (and the pipelines connected to them) by the
out-of-order engine. The pipeline that takes the largest number of
cycles to execute all its instructions determines the runtime.  The
worst-case assumption would be an execution fully determined by the
critical path through the code, heeding all dependencies.  In
practice, the actual runtime will be between these limits unless other
bottlenecks apply that are not covered by the in-core execution, such
as data transfers from beyond the L1 cache, instruction cache misses,
etc. Even if a loop body contains strong dependencies the throughput
assumption may still hold if there are no loop-carried dependencies.

\begin{table*}[tb]
\centering
\caption{\label{tab:instructions}Measured worst-case latency and inverse throughput for floating-point arithmetic instructions.
For all of these numbers, lower is better.}
\begin{tabular}{ l @{\hskip 0.2in} c c c c @{\hskip 0.1in} c c c c }
    \toprule
                                & \multicolumn{4}{c}{Latency [cy]}     &                 \multicolumn{4}{c}{\makebox[0pt][c]{Inverse throughput [cy/inst.]}}        \\
    \midrule
    \textmu{}arch               & BDW           & HSW           & IVB           & SNB            & BDW               & HSW       & IVB       & SNB   \\
    \midrule
    \texttt{vdivpd} (AVX)       & 24            & 35            & 35            & 45            & 16                & 28        & 28        & 44        \\
    \texttt{divpd} (SSE)        & 14            & 20            & 20            & 22            & 8                 & 14        & 14        & 22        \\
    \texttt{divsd} (scalar)     & 14            & 20            & 20            & 22            & \textbf{4.5}               & 14        & 14        & 22        \\
    \texttt{vdivps} (AVX)       & 17            & 21            & 21            & 29            & 10                & 14        & 14        & 28        \\
    \texttt{divps} (SSE)        & 11            & 13            & 13            & 14            & 5                 & 7         & 7         & 14        \\
    \texttt{divss} (scalar)     & 11            & 13            & 13            & 14            & \textbf{2.5}               & 7         & 7         & 14        \\
    \midrule
    \texttt{vsqrtpd} (AVX)      & 35            & 35            & 35            & 44            & 28                & 28        & 28        & 43        \\
    \texttt{sqrtpd} (SSE)       & 20            & 20            & 20            & 23            & 14                & 14        & 14        & 22        \\
    \texttt{sqrtsd} (scalar)    & 20            & 20            & 20            & 23            & \textbf{7}                 & 14        & 14        & 22        \\
    \texttt{vsqrtps} (AVX)      & 21            & 21            & 21            & 23            & 14                & 14        & 14        & 22        \\
    \texttt{sqrtps} (SSE)       & 13            & 13            & 13            & 15            & 7                 & 7        & 7         & 14        \\
    \texttt{sqrtss} (scalar)    & 13            & 13            & 13            & 15            & \textbf{4}                 & 7        & 7         & 14        \\
    \midrule
    \texttt{vrcpps} (AVX)       & 7             & 7             & 7             & 7             & 2                 & 2         & 2         & 2   \\
    \texttt{rcpps} (SSE, scalar)& 5             & 5             & 5             & 5             & 1                 & 1         & 1         & 1   \\
    \midrule
    \texttt{*add*}                & 3,4\textsuperscript{\textdagger} 
                                                & 3             & 3             & 3             & 1                 & 1         & 1         & 1      \\
    \texttt{*mul*}                & 3             & 5             & 5           & 5             & 0.5               & 0.5       & 1         & 1      \\
    \texttt{*fma*}                & 5,6\textsuperscript{\textdaggerdbl}
                                                & 5,6\textsuperscript{\textsection}
                                                                & ---           & ---           & 0.5               & 0.5       & ---       & ---   \\
    \bottomrule
    \multicolumn{9}{l}{\textsuperscript{\textdagger}\footnotesize{SP/DP AVX addition: 3 cycles; SP/DP SSE and scalar addition: 4 cycles}} \\
    \multicolumn{9}{l}{\textsuperscript{\textdaggerdbl}\footnotesize{SP/DP AVX FMA: 5 cycles; SP/DP SSE and scalar FMA: 6 cycles}} \\
    \multicolumn{9}{l}{\textsuperscript{\textsection}\footnotesize{SP scalar FMA: 6 cycles; all other: 5 cycles}}
\end{tabular}
\end{table*}
Calculating the throughput and critical path predictions requires
information about the maximum throughput and latency of all relevant
instructions as well as general limits such as decoder/retirement
throughput, L1I bandwidth, and the number and types of address
generation units. The Intel Architecture Code
Analyzer\footnote{\url{http://software.intel.com/en-us/articles/intel-architecture-code-analyzer/}} 
(IACA) can help with this, but it is proprietary
software with an unclear future development path and it does not
always yield accurate predictions. Moreover, it can only analyze
object code and does not work on the high-level language
constructs. Thus one must often revert to manual analysis to get
predictions for the best possible code, even if the compiler cannot
produce it. In Table~\ref{tab:instructions} we give worst-case measured latency
and inverse throughput numbers for arithmetic instructions
in AVX, SSE, and scalar mode. In the following we point out some
notable changes over the four processor generations.

The most profound change happened in the performance of the divide
units. From SNB to BDW we observe a massive decrease in latency and an
almost three-fold increase in throughput for AVX and SSE instructions,
in single and double precision alike. Divides are still slow compared
to multiply and add instructions, of course. The fact that the 
divide throughput \emph{per operation} is the same for AVX and SSE
is well known, but with BDW we see a significant rise in scalar
divide throughput, even beyond the documented limit of one
instruction every five cycles. The scalar square root instruction shows
a similar improvement, but is in line with the documentation.

The standard multiply, add, and fused multiply-add instructions have
not changed dramatically over four generations, with two exceptions:
Together with the introduction of FMA instructions with HSW, it became
possible to execute two plain multiply (but not add) instructions per
cycle. The latency of the add instruction in scalar and SSE mode on
BDW has increased from three to four cycles; this result 
is not documented by Intel for BDW but announced for AVX code
in the upcoming Skylake architecture. The fma instruction shows
the same characteristic (latency increase from 5 to 6 cycles 
when using SSE or scalar mode).

One architectural feature that is not directly evident from
single-instruction measurements is the number of address generation
units (AGUs). Up to IVB there are two such units, each paired with a
LOAD unit with which it shares a port. As a consequence, only two
addresses per cycle can be generated. HSW introduced a third AGU on
the new port 7, but it can only handle simple addresses for STORE
instructions, which may lead to some restrictions. See Sect.~\ref{sec:L1}
for details.



%
%
\subsection{L1 cache/AGU}\label{sec:L1}

The cores of all four microarchitectures feature two load units and one store
unit. The data paths between each unit and the L1 cache are 16\,B on SNB and
IVB, and 32\,B on HSW and BDW. The theoretical bandwidth is thus
48\,B/cy on SNB and IVB and 96\,B/cy on HSW and BDW; however, several
restrictions apply.

\begin{figure}[tb]
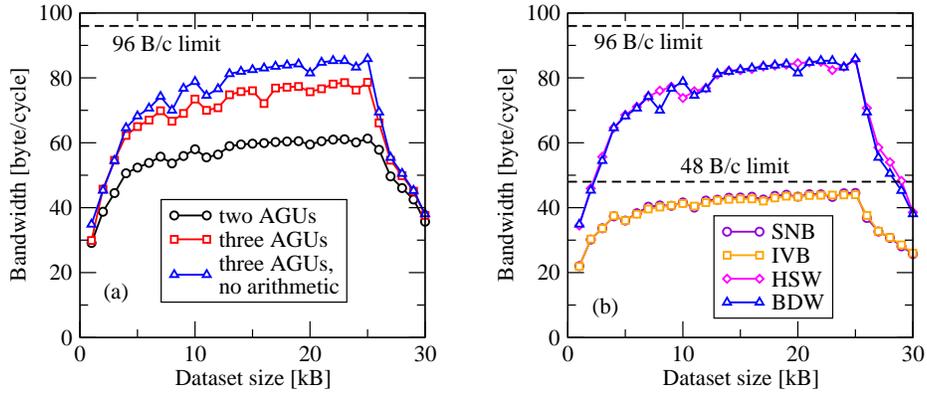

    \centering
    \includegraphics*[scale=0.5]{L1-BW-bdw}\hfill
    \includegraphics*[scale=0.5]{L1-BW-arch-comp}
    \caption{\label{fig:L1-BW}(a) L1 bandwidth achieved with STREAM triad and various optimizations on BDW. (b) Comparison of achieved L1 bandwidths using STREAM triad on all microarchitectures.}
\end{figure}

An \ac{AVX} vectorized STREAM triad benchmark uses two AVX loads, one AVX
\ac{FMA}, and one AVX store instruction to update four DP elements. On HSW and
BDW, only two \acp{AGU} are capable of performing the necessary address
computations, i.e., (base + scaled index + offset), typically
used in streaming memory accesses; HSW's newly introduced third store \ac{AGU}
can only perform offset computations. This means that only two
addresses per cycle can be calculated, limiting the
L1 bandwidth to 64\,B/cy. STREAM triad performance using only two \acp{AGU} is
shown in Figure~\ref{fig:L1-BW}a.  One can make use of the
new \ac{AGU} by using one of the ``fast LEA'' units (which can perform only
indexed and no offset addressing) to pre-compute an intermediate address, which
is then used by the simple \ac{AGU} to complete the address calculation.
This way both
AVX load units and the AVX store unit can be used simultaneously. When the
store is paired with address generation on the new store AGU, both micro-ops
are fused into a single micro-op. This means that the four micro-op per cycle
front end retirement constraint should not be a problem: in each cycle two AVX
load instructions, the micro-op fused AVX store instruction, and one AVX
\ac{FMA} instruction is retired. With sufficient unrolling, loop instruction
overhead becomes negligible and the bandwidth should approach 96\,B/cy.
Figure~\ref{fig:L1-BW} shows, however, that micro-op throughput still seems to be
the bottleneck because bandwidth can be further increased by removing
the \ac{FMA} instructions from the loop body.

Figure~\ref{fig:L1-BW}b compares the bandwidths achievable by different
microarchitectures (using no arithmetic instructions on HSW and BDW
for the reasons described above). On SNB and IVB a regular
STREAM triad code can almost reach maximum theoretical L1 performance
because it only requires half the number of address calculations per cycle,
i.e., two \acp{AGU} are sufficient to generate three
addresses every two cycles.

\subsection{Gather}\label{sec:gather}

Vector gather is a microcode solution for loading noncontinuous data into vector
registers. The instruction was first implemented in Intel multicore CPUs
with AVX2 on HSW. The first implementation offered a
poor latency (i.e., the time until all data was placed in the vector register)
and using hand-written assembly to manually load distributed data into vector
registers proved to be faster than using the gather instruction in some cases
\cite{Hofmann:2014}.

\begin{table*}[tb]
\centering
\caption{\label{tab:gather_latency}Time in cycles per gather instruction on HSW and BDW depending on data distribution across CLs.}
\begin{tabular}{ c @{\hskip 0.2in} c c c c @{\hskip 0.2in} c c c c }
    \toprule
    Microarchitecture           & \multicolumn{4}{c}{Haswell-EP}                    & \multicolumn{4}{c}{Broadwell-EP}      \\
    \midrule
     Location of data           & L1        & L2        & L3        & Mem           & L1        & L2    &  L3       & Mem   \\
    \midrule
    Distributed across 1 CLs    & 12.3      & 12.3      & 12.4      & 15.5          & 7.3       & 7.3   &  7.7      & 13.3  \\
    Distributed across 2 CLs    & 12.5      & 12.5      & 13.2      & 23.0          & 7.5       & 7.6   &  11.0     & 24.5  \\
    Distributed across 4 CLs    & 12.5      & 12.7      & 20.6      & 42.7          & 7.5       & 9.9   &  20.0     & 47.5  \\
    Distributed across 8 CLs    & 12.3      & 18.4      & 38.5      & 89.3          & 7.3       & 18.1  &  38.2     & 94.4  \\
    \bottomrule
\end{tabular}
\end{table*}
Table~\ref{tab:gather_latency} shows the gather instruction latency for both
HSW and BDW. The latency depends on where the data  is
coming from and, in case data is not in L1, over how many \acp{CL}
it is distributed.  We find that the instruction is 40\% faster on BDW in L1.
When data is coming from L2 on HSW and distributed across eight \acp{CL}, the
latency is dominated by time required to transfer eight \acp{CL}
from L2 to L1 cache. On BDW, this effect is already visible when data is coming
from the L2 cache and distributed across four \acp{CL}. BDW's improvement of the
instruction offers no returns when the latency is dominated by \ac{CL}
transfers, which is the case when loading more than four \acp{CL} from L2, two
from L3, or one from memory.

\section{L2 cache}

According to official documentation, the L2 cache bandwidth on HSW
was increased from 32\,B/cy to 64\,B/cy compared to IVB.
To validate this expectation, knowledge about overlapping
transfers in the cache hierarchy is required. The ECM model for x86
assumes that no \acp{CL} are transferred between L2 and L1 in any cycle
in which a LOAD instruction retires. Hence, the maximum of 64\,B/cy
can never be attained by design but an improvement may still be expected.
To derive the time spent
transferring data, cycles in which load instructions are
retired are subtracted from the overall runtime with an in-L2 working set.
The resulting bandwidth should be compared with the documented theoretical
maximum.


\begin{table*}[b]
\centering
\caption{\label{tab:L2_BW}Measured L1-L2 bandwidth on different microarchitectures for dot product and STREAM triad access patterns.}
\begin{tabular}{ l @{\hskip 0.1in} l @{\hskip 0.1in} c @{\hskip 0.1in} c @{\hskip 0.1in} c @{\hskip 0.1in} c }
    \toprule
    Pattern         & Code                      &   SNB             & IVB               & HSW               & BDW   \\
    \midrule
    Dot product     & \texttt{dot+=A[i]+B[i]}   & 28   & 27   & 43   & 43 \\
    STREAM triad    & \texttt{A[i]=B[i]+s*C[i]} & 29  & 29  & 32  & 32 \\
    \bottomrule
\end{tabular}
\end{table*}
Table~\ref{tab:L2_BW} shows the measured bandwidths for a dot product (a
load-only benchmark) and the STREAM triad. Both SNB and IVB operate near the
specified bandwidth of 32\,B/cy for both access patterns.  Although HSW and BDW
offer bandwidth improvements, especially in case of the dot product,
measured bandwidths are significantly below the advertised 64\,B/cy.

The question arises of how this result may be incorporated into the ECM model.
Preliminary experiments indicate that the ECM predictions for in-L3 data
are quite accurate when assuming theoretical L2 throughput. We could thus
interpret the low L2 performance as a consequence of a latency penalty,
which can be overlapped when the data is further out in the hierarchy.
Further experiments are needed to substantiate this conjecture.

\section{Uncore}


\subsection{L3 cache}
\subsubsection{Cluster-on-Die}
\label{sec:cod}

Together with the dual-ring interconnect, HSW introduced the \ac{CoD} mode, in
which a single chip can be partitioned into two equally-sized \ac{NUMA} clusters.
HSW features a so-called ``eight plus $x$'' design, in
which the first physical ring features eight cores and the second ring contains
the remaining cores (six for our HSW chip).  This asymmetry leads to a scenario
in which the seven cores in the first cluster domain are physically located on
the first ring; the second cluster domain contains the remaining core from the
first and six cores from the second physical ring. The asymmetry was removed on
BDW: here both physical rings are of equal size so both cluster domains contain
cores from dedicated rings.
\ac{CoD} is intended for \ac{NUMA}-optimized code and impacts L3 scalability
and latency and, implicitly, main memory bandwidth because it
uses a dedicated snoop mode that makes use of a directory to avoid unnecessary
snoop requests (see Section~\ref{sec:snoop} for more details).
\begin{figure}[!tb]
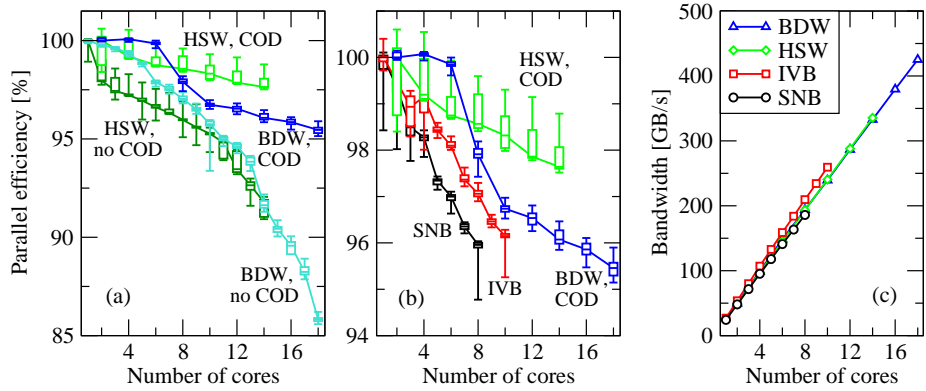

    \centering
    \includegraphics*[scale=0.5]{arch-comp-L3-BDWHSW}
    \includegraphics*[scale=0.5]{arch-comp-L3-scaling}\hfill
    \includegraphics*[scale=0.5]{arch-comp-L3-BW}
    \caption{\label{fig:L3-BW-and-scalability}(a) L3 scalability on
      HSW and BDW depending on whether \ac{CoD} is used. (b)
      Comparison of microarchitectures regarding L3 scalability.
      (c) Absolute L3 bandwidth for STREAM triad as function of cores
      on different microarchitectures.}
\end{figure}

Figure~\ref{fig:L3-BW-and-scalability}a shows the influence of \ac{CoD} on
L3 bandwidth (using STREAM triad) for HSW and BDW. When data is distributed
across both rings on HSW, the
parallel efficiency of the L3 cache is 92\%; it can be raised to 98\% by using
\ac{CoD}. The higher core count of BDW results in a more pronounced effect;
here, parallel efficiency is only 86\% in non-CoD mode.
Using \ac{CoD} the efficiency goes above 95\%.
Figure~\ref{fig:L3-BW-and-scalability}b shows that HSW and
BDW with CoD offer similar L3 scalability as SNB and IVB.

Assuming an $n$-core chip, the topological diameter (and with it the average
distance from a core to data in an L3 segment) is smaller in each of the
$n/2$-core cluster domains in comparison to the non-\ac{CoD} domain consisting
of $n$ cores. Shorter ways between cores and data result in lower latencies
when using \ac{CoD} mode. On BDW, the L3 latency is 41 cycles when with
\ac{CoD} and 47 cycles without (see Table~\ref{tab:mem_latency}).

\subsection{Memory}

\subsubsection{Snoop modes}
\label{sec:snoop}

Starting with HSW, the QPI
snoop mode can be selected at boot time.
HSW supports three snoop modes: \ac{ES}, \ac{HS}, and
\ac{DIR} (often only indirectly selectable by enabling \ac{CoD} in the BIOS)
\cite{Molka:2015,Intel:registers,Intel:OSB}.  BDW introduced a fourth snoop mode
called \ac{HS}+\ac{OSB} \cite{Intel:OSB}.  The remainder of this section
discusses the differences among the modes and the immediate impact
on memory latency and bandwidth.

On a L3 miss inside a \ac{NUMA} domain, in addition to fetching the \ac{CL}
containing the requested data from main memory, cache coherency mandates
other \ac{NUMA} domains be checked for modified copies of the \ac{CL}.
Attached to each L3 segment is a \ac{CA} responsible for sending
and receiving snoop information. In addition to multiple \acp{CA},
each \ac{NUMA} domain features a \ac{HA}, which plays a major role in snooping.

In \ac{ES}, snoop requests are sent directly from the \ac{CA} of the L3 segment
in which the L3 miss occurred to the respective\footnote{\acp{CL} are mapped to
L3 segments based on their addresses according to a hashing function. Thus,
each \ac{CA} knows which \ac{CA} in other \ac{NUMA} domains is responsible for a
certain \ac{CL}.} \acp{CA} in other \ac{NUMA} domains. Queried remote \acp{CA}
directly respond back to the requesting \ac{CA}; in addition, they report to
the \ac{HA} in the requesting \ac{CA}'s domain, so it can resolve potential
conflicts. \ac{ES} involves a lot of requests and replies, but offers low
latencies.

In \ac{HS}, \acp{CA} forward snoop requests to their \ac{NUMA} domain's \ac{HA}. The
\ac{HA} proceeds to fetch the requested \ac{CL} from memory but stalls snoop
requests to remote \ac{NUMA} domains until the \ac{CL} is available. 
For each \ac{CL}, so-called directory information is stored
in its memory ECC bits.  The bits indicate whether a copy of the \ac{CL} exists in
other \ac{NUMA} domains. The directory bits only tell whether a \ac{CL} is
present or not in other \ac{NUMA} domains; they do not tell which \ac{NUMA}
domain to query, so snoops have to broadcast to all \ac{NUMA} domains. By waiting for
directory data, unnecessary snoop requests are avoided at the cost of higher
latency due to delayed snoops. By reducing snoop requests, overall bandwidth
can be increased.  As in \ac{ES}, potentially queried remote \acp{CA} respond
to the initiating \ac{CA} and \ac{HA}, which resolves potential conflicts.

In \ac{DIR}, a two-step approach is used. Starting with HSW, each \ac{HA}
features a 14\,kB directory cache (also called ``HitMe'' cache) holding
additional directory information for \acp{CL} present in remote \ac{NUMA}
domains.\footnote{Investigations using the \texttt{HITME\_*} performance
counter events indicate this cache is exclusively used in \ac{DIR} mode.} In
addition to the directory information recorded in the ECC bits, the directory
cache stores the particular \ac{NUMA} domain in which the copy of a CL
resides; this means that on a hit in the directory cache only a single snoop
request has to be sent. This mechanism further reduces snoop traffic,
potentially increasing bandwidth.  When the directory cache is hit, latency is
also improved in \ac{DIR} compared to \ac{HS}, because snoops are not delayed
until directory information stored in ECC bits from main memory becomes
available.  In case of a directory cache miss, \ac{DIR} mode proceeds similarly
to \ac{HS}. Note, however, that \ac{DIR} mode is recommended only for
\ac{NUMA}-aware workloads. The directory cache can only hold data for a small
number of \acp{CL}. If the number of \acp{CL} shared between both cluster
domains exceeds the directory cache capacity , \ac{DIR} mode degrades to
\ac{HS} mode, resulting in high latencies.

BDW's new \ac{HS}+\ac{OSB} mode works similarly to \ac{HS}. However, \acp{HA}
will send opportunistic snoop requests while waiting for directory information
stored in the ECC bits under ``light'' traffic conditions. Latency is reduced
in case the directory information indicates snoop requests have to be sent,
because they were already sent opportunistically. Redundant snoop
requests are not supposed to impact performance under ``light'' traffic
conditions.
\begin{table*}[tb]
\centering
\caption{\label{tab:mem_latency}Measured access latencies of all memory hierarchy levels in base frequency core cycles}
\begin{tabular}{ c @{\hskip 0.4in} c @{\hskip 0.4in} c @{\hskip 0.4in} c @{\hskip 0.4in} c}
    \toprule
    \textmu{}arch           & L1        & L2        & L3                    & MEM   \\
    \midrule
    SNB                         & 4         & 12        & 40                    & 230   \\ 
    IVB                         & 4         & 12        & 40                    & 208   \\ 
    HSW                         & 4         & 12        & 37\textsuperscript{2}  & 168\textsuperscript{6}\\
    BDW                         & 4         & 12        & 47\textsuperscript{1}, 41\textsuperscript{2}  & 248\textsuperscript{3}, 280\textsuperscript{4}, 190\textsuperscript{5}, 178\textsuperscript{6} \\
    \bottomrule
    \multicolumn{5}{l}{\textsuperscript{1}\footnotesize{COD disabled}, \textsuperscript{2}\footnotesize{COD enabled},\textsuperscript{3}\footnotesize{ES}, \textsuperscript{4}\footnotesize{HS}, \textsuperscript{5}\footnotesize{HS+OSB}, \textsuperscript{6}\footnotesize{DIR}}
\end{tabular}
\end{table*}

The impact of snoop modes is largest on main memory latency. As expected,
\ac{DIR} produces the best results with 178\,cy (see
Table~\ref{tab:mem_latency}).  Pointer chasing in main memory does not generate
a lot of traffic on the ring interconnect, which is why \ac{HS}+\ac{OSB} will
generate opportunistic snoops, achieving a latency of 190\,cy.  The
difference in latency of 12\,cy compared to \ac{DIR} can be explained
through shorter paths inside a single cluster domain in \ac{CoD} mode.  We
measured an L3 latency of 41\,cy for \ac{CoD} and 47\,cy for
non-\ac{CoD} mode.  Since memory accesses pass through the interconnect twice
(one to request the \ac{CL}, once to deliver it) the memory latency of
non-\ac{CoD} mode is expected to be twice the L3 latency penalty of six cycles. In
\ac{ES}, the requesting \ac{CA} has to wait for its \ac{HA} to acknowledge that it
received all snoop replies from the remote \acp{CA}, which causes a latency
penalty.  On BDW, the measured memory latency is 248\,cy. As expected,
\ac{HS} offers the worst latency at 280\,cy, because necessary snoop
broadcasts are delayed until directory information becomes available from main
memory.
\begin{figure}[!tb]
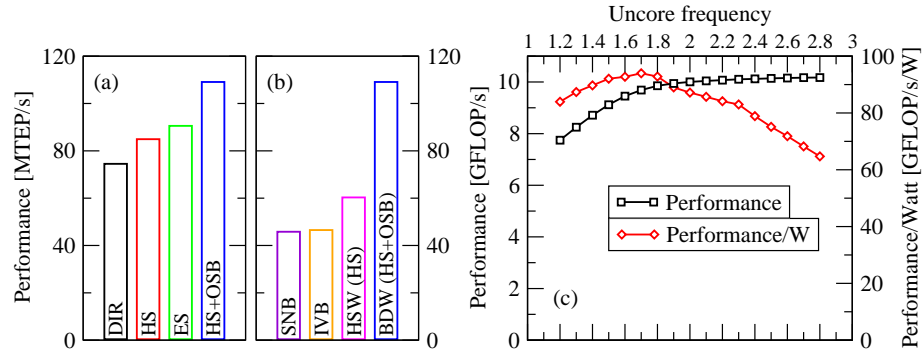

    \centering
    \includegraphics*[scale=0.5]{graph500-bar-BDW}
    \includegraphics*[scale=0.5]{graph500-bar-comp}
    \includegraphics*[scale=0.5]{HPCG-UFS}
    \caption{\label{fig:graph500_and_HPCG_UFS_perf}(a) Graph500 performance in millions of traversed edges per second (MTEP/s) as function of snoop mode on BDW. (b) Graph500 performance of all chips. (c) HPCG performance and performance per Watt as function of Uncore frequency.}
\end{figure}

Graph500 was chosen to evaluate the influence of snoop modes on the performance of
latency-sensitive workloads. Figure~\ref{fig:graph500_and_HPCG_UFS_perf}a shows Graph500
performance for a single BDW chip.
A direct correlation between latency and performance can be observed for HS,
ES, and HS+OSB. DIR mode performs worst despite offering the best memory latency.  This
can be explained by the non-\ac{NUMA}-awareness of the Graph500 benchmarks.  Too
much data is shared between both cluster domains; this means the directory
cache can not hold information on all shared \acp{CL}. As a result, snoops are
delayed until directory information from main memory becomes available.
Figure~\ref{fig:graph500_and_HPCG_UFS_perf}b shows an overview of Graph500 performance
on all chips and the qualitative improvement offered by the new
\ac{HS}+\ac{OSB} snoop mode introduced with BDW.
\begin{figure}[!tb]
    \centering
    \includegraphics*[width=0.9\textwidth]{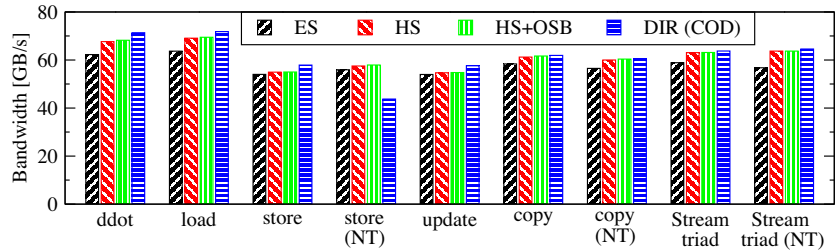}
    \caption{\label{fig:mem_bw_broadep2}Sustained main memory bandwidth on BDW for various access patterns. NT=nontemporal stores.}
\end{figure}

\begin{figure}[!tb]
    \centering
    \includegraphics*[width=0.9\textwidth]{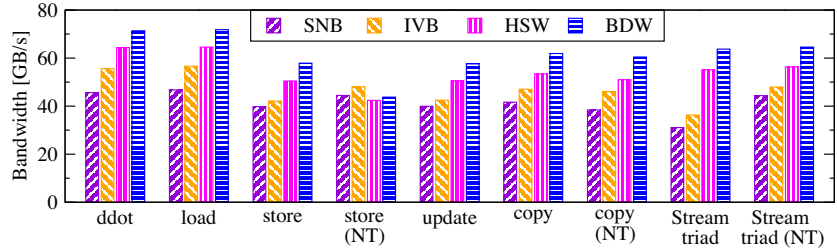}
    \caption{\label{fig:mem_bw_evolution}Comparison of sustained main memory bandwidth across microarchitectures for various access patterns.}
\end{figure}

The effect of snoop mode on memory bandwidth for BDW is shown in
Fig.~\ref{fig:mem_bw_broadep2}.  The data is roughly in line
with the reasoning above. For \ac{NUMA}-aware workloads, \ac{DIR} should produce the
least snoop traffic due to snoop information stored in the directory
cache.  This is reflected in a slightly better bandwidth compared to other
snoop modes (with the exception of the \ac{NT} store access pattern, which
seems to be a toxic case for \ac{DIR} mode).  \ac{DIR} offers up to 10\,GB/s
more for load-only access patterns when compared to \ac{ES}, which produces the
most amount of snoop traffic. The effect is less pronounced but still
observable when comparing \ac{DIR} to \ac{HS} and \ac{HS}+\ac{OSB}.
Figure~\ref{fig:mem_bw_evolution} shows the evolution of sustained memory
bandwidth for all examined microarchitectures, using the best snoop mode on
HSW and BDW. Increases in bandwidths
over the generations is explained by new DDR standards as well as increased
memory clock speeds (see Table~\ref{tab:testbed}).

\subsection{Uncore clock, bandwidth, and energy efficiency}
\label{sec:ufs}

Before HSW, the Uncore was clocked at the same frequency as the cores. Starting
with HSW, the Uncore has its own clock frequency. The motivation for
this lies in potential energy savings: When cores do not require much
data via the Uncore (i.e., from/to L3 cache and main memory) the
Uncore can be slowed down to save power. This mode of operation is called
\ac{UFS}. For our BDW chip, the Uncore frequency can vary automatically between
1.2\,--2.8\,GHz, but one can also define custom minimum and
maximum settings within this range via \acp{MSR}.

We examine the default \ac{UFS} behavior for both extremes of the Roof{}line
spectrum and use HPCG as a bandwidth-bound and LINPACK as a compute-bound
benchmark. Our findings indicate that at both ends of the spectrum, \ac{UFS}
tends to select higher than necessary frequencies, pointlessly boosting
power and in the case of LINPACK even hurting performance.

Figure~\ref{fig:graph500_and_HPCG_UFS_perf}c shows HPCG performance and
energy efficiency versus Uncore frequency
for a fixed core clock of 2.3\,GHz on HSW. We find that the
Uncore is the performance bottleneck only for Uncore frequencies below
2.0\,GHz.  Increasing it beyond this point does not
improve performance, because main memory is now the bottleneck.
Using performance counters the Uncore frequency was determined to be the
maximum of 2.8\,GHz when running HPCG in \ac{UFS} mode.  The energy efficiency of
64.7\,GFLOP/s/W at 2.8\,GHz is 26\% lower than the 87.2\,GFLOP/s/W observed at
2.0\,GHz Uncore frequency, at almost the same performance. Energy efficiency
can be increased even more by further lowering the Uncore
clock; however, below 2.0\,GHz performance is degraded.
\begin{figure}[!tb]
    \centering
    \includegraphics*[width=0.9\textwidth]{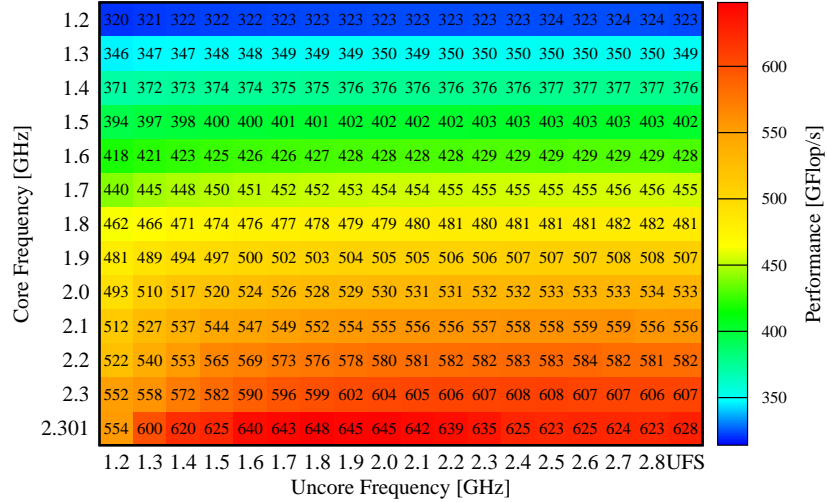}
    \caption{\label{fig:HPL_perf_heatmap}LINPACK performance on BDW as a function of core and uncore frequency.}
\end{figure}

For LINPACK, we observe a particularly interesting side effect of varying
Uncore frequency. Figure~\ref{fig:HPL_perf_heatmap} shows LINPACK performance
on BDW as a function of core and Uncore clock. Note that in Turbo mode,
the performance increases when going from the highest Uncore frequencies
towards 1.8\,GHz. This effect is caused by Uncore
and cores competing for the chip's TDP. When the Uncore clovk speed is reduced, a
larger part of the chip's power budget can be consumed by the cores, which
in turn boost their frequency. The core frequency in Turbo mode is 2479\,MHz
when the Uncore clock is set to 2.8\,GHz (the Uncore actually only achieves a
clock rate of 2475\,MHz) vs 2595\,MHz when the Uncore clock is set to 1.8\,GHz.
Below 1.8\,GHz the
CPU frequency increases further, e.g., to 2617\,MHz at an Uncore clock of
1.7\,GHz and up to 2720\,MHz at an Uncore clock of 1.2\,GHz. LINPACK performance
starts to degrade at this point despite an increasing core frequency
due to the Uncore becoming a data bottleneck.
In \ac{UFS} mode, the Uncore
is clocked at 2489\,MHz and the cores run at 2491\,MHz.  Compared to
the optimum, \ac{UFS} degrades performance by 3\%.
Energy efficiency is reduced by 6\% from 4.94\,GFLOP/s/W at an Uncore clock of
1.8\,GHz to 4.65\,GFLOP/s/W in \ac{UFS}. The most energy-efficient operating
point for LINPACK is 5.74\,GFLOP/s/W at a core clock of of 1.6\,GHz and an
Uncore clock of 1.2\,GHz.

\section{Conclusions and outlook}

We have conducted an analysis of core- and chip-level performance features of four
recent Intel server CPU architectures.
Previous findings about the behavior of clock speed and
its interaction with thermal design limits on Haswell and Broadwell CPUs could
be confirmed. Overall the documented instruction latency and
throughput numbers fit our measurements, with slight deviations
in scalar DP divide throughput and SSE/scalar add and fused
multiply-add latency on Broadwell. We could also demonstrate the
consequences of limited instruction throughput and the special
properties of Haswell's and Broadwell's address generation units for
L1 cache bandwidth.

Our microbenchmark results have unveiled that the gather instruction,
which was newly introduced with the AVX2 instruction set, was finally
implemented on Broadwell in a way that makes it faster than hand-crafted
assembly. The L2 cache on Haswell and Broadwell does not keep its
promise of doubled bandwidth to L1 but only delivers between 32 and
43\,B/cy, as opposed to Sandy Bridge and Ivy Bridge, which
get close to their architectural limit of 32\,B/cy.

The scalable L3 cache was one of the major innovations in the
Sandy Bridge architecture.
On Haswell and Broadwell, the bandwidth scalability of the L3
cache is substantially improved in Cluster on Die (CoD) mode.
Even without CoD the full-chip efficiency (at up to 18 cores)
is never worse than 85\%. In the memory domain we find, unsurprisingly,
that CoD provides the lowest latency and highest memory bandwidth
(except with streaming stores), but the irregular Graph500
benchmark shows a 50\% speedup on Broadwell when switching to
non-CoD and Home Snoop with Opportunistic Snoop Broadcast.

Finally, our analysis of core and Uncore clock speed domains
has exhibited significant potential for saving energy
in a sensible setting of the Uncore frequency, without
sacrificing execution performance.

Future work will include a thorough evaluation of the ECM performance
model on all recent Intel architectures, putting to use the insights
generated in this study. Additionally, existing analytic power and
energy consumption models will be extended to account for the Uncore
power more accurately. Significant changes in performance and power
behavior are expected for the upcoming Skylake architecture, such as
(among others) an L3 victim cache and AVX-512 on selected models,
and will pose challenges of their own.

\bibliographystyle{splncs03}


\end{document}